\documentclass[showpacs,twocolumn,aps,prb,amsmath,amssymb,superscriptaddress]{revtex4}
\usepackage{graphicx}
\usepackage{dcolumn}
\usepackage{bm}
\usepackage{color}

\begin{document}

\baselineskip=0.5cm
\renewcommand{\thefigure}{\arabic{figure}}
\title{Spin polarization dependence of quasiparticle properties in graphene}
\author{Alireza Qaiumzadeh}
\affiliation{Department of Physics, Norwegian University of Science and Technology, NO-7491 Trondheim, Norway }
\author{Kh. Jahanbani}
\affiliation{School of Physics, Institute for Research in
Fundamental Sciences (IPM), Tehran 19395-5531, Iran}
\affiliation{Institute for Advanced Studies in Basic Sciences (IASBS), Zanjan,
P. O. Box 45195-1159, Iran}
\author{Reza Asgari}
\email{asgari@ipm.ir} \affiliation{School of Physics, Institute for Research in
Fundamental Sciences (IPM), Tehran 19395-5531, Iran}

\begin{abstract}
We address spin polarization dependence of graphene's Fermi liquid properties
quantitatively using a microscopic Random Phase Approximation theory in an interacting spin-polarized Dirac electron system. We show an enhancement of the minority-spin many-body velocity renormalization at fully spin polarization due to reduction in the electron density and consequently increase in the interaction between electrons near the Fermi surface. We also show that the spin dependence of the Fermi velocity in the chiral Fermi systems is different than that in a conventional two-dimensional electron liquid. In addition, we show that the ratio of the majority- to minority-spin lifetime is smaller than unity and related directly
to the polarization and electron energy. The spin-polarization dependence of the carrier Fermi velocity is of significance in various spintronic applications.

\end{abstract}
\maketitle

\section{Introduction}

Graphene is a two-dimensional crystal of carbon atoms has been recently
discovered~\cite{novoselov}. This stable
crystal has attracted considerable attention~\cite{allan}
because of its unusual effective many-body properties~\cite{yafis,
polini, polini2,ee interaction, asgari_im, kotov_arXiv_2010} that follow from chiral band states and
because of potential applications. The low energy
quasiparticle excitations energy in graphene are linearly
dispersing, described by Dirac cones at the edges of the first
Brillouin zone.

Stable non-reactive graphene layers on top of ferromagnetic materials
~\cite{dedkov} might be used as sources of spin-polarized electrons. Electron sources are used in all domains ranging from technical devices like cathode-ray tubes to large scale scientific experiments like electron accelerators. This is of great interest
for studies of magnetic systems in condensed matter physics, including the field of spintronics.

Graphene's spin-transport properties are expected to be particularly interesting, with predictions for extremely long coherence times and intrinsic spin-polarized states at zero field~\cite{yazyev}. Spin-polarized electron emission from the graphene/{\rm Ni} system before and after
exposure to oxygen has been recently studied~\cite{dedkov2} and the spin polarization of secondary electrons obtained from this system upon photoemission and suggested to use such
passivated {\rm Ni} surface as a source of spin-polarized electrons stable against adsorption of reactive gases. The resolve spin transport from conductance features that are caused by quantum interference has already been measured~\cite{lundeberg}. These features split visibly in an in-plane magnetic field, similar to Zeeman splitting in atomic and quantum-dot systems. As a result, spin-up and spin-down conductance contributions at finite field are offset in gate voltage, leading to Zeeman splitting of interference features in a gate-voltage~\cite{lundeberg}.

Many electronic and optical properties of graphene could be explained within a single-particle picture in which electron-electron interactions are completely neglected. The discovery of the fractional quantum Hall effect in graphene~\cite{fqhe} represents an important hallmark in this context. By now there is a large body of experimental work~\cite{eeinteractionsgraphene,bostwick_science_2010,kotov_arXiv_2010} showing the relevance of electron-electron interactions in a number of key properties of graphene samples of sufficiently high quality.

Conventional two-dimensional electron gas (2DEG), on the other hand, has been a
fertile source of surprising new physics for more than four decades.
Although graphene was only isolated for the first time in 2004 and it is still at an early stage, it is already clear~\cite{novoselov2} that the strong field
properties of Dirac electrons in graphene are different from
and as rich as those of a semiconductor heterojunction 2DEG. The
Fermi liquid phenomenology of Dirac electrons in
graphene~\cite{polini, polini2} and conventional
2DEG~\cite{asgari2} have the same structure, since both systems are
isotropic and have a single circular Fermi surface. The strength of
interaction effects in a conventional 2DEG increases with decreasing
carrier density. At low densities, the quasiparticle weight $Z$ is
small, the velocity is suppressed~\cite{asgari2}, the charge
compressibility changes sign from positive to negative,
and the spin-susceptibility is strongly enhanced~\cite{asgari3}.
These effects emerge from an interplay between exchange interactions
and quantum fluctuations of charge and spin in the 2DEG.

In addition, effective mass or the effective Fermi velocity
is an important concept in Landau's Fermi liquid theory since it
provides a direct measure of the many-body interactions in the
electron system. In the highly interacting, dilute, paramagnetic regime in the 2DEG
the effective Fermi velocity, which is defined by the effective mass as $v^*=\hbar k_{\rm F}/m^*$, is significantly diminished compared to its band
value and tends to decrease with increasing
$r_{s}$~\cite{asgari2,asgari3,PadmanabhanPRL08,AsgariSSC04}, the so-called Wigner-Seitz radius. Recent
measurements of the effective mass for two-dimension electrons confined to AlAs quantum
wells revealed that, when the 2DES is fully valley- and
spin-polarized, the effective mass is suppressed down to values near or
even slightly below the band mass\cite{PadmanabhanPRL08,GangadharaiahPRL05, GokmenPRL08,GokmenUNP09}. A sophisticate theoretical calculation has been shown~\cite{asgari-shayegan} that in an interacting, fully spin-polarized 2DES the absence (freezing out) of spin fluctuations reduces the effective mass below its band value, in agreement with experimental data. Furthermore, the spin-up and spin-down effective masses
from magnetotransport measurements at different temperatures for a 2DEG and the effective hole mass measurements through analyzing the temperature dependence of
Shubnikov-de Haas oscillations in dilute 2D hole systems have been recently reported~\cite{wei}.

In the Dirac electrons in graphene, it was
shown~\cite{yafis,polini,polini2, elias} that interaction effects also
become noticeable with decreasing density that the
quasiparticle weight $Z$ tends to larger values, that the velocity
is enhanced rather than suppressed, and that the influence of
interactions on the compressibility and the spin-susceptibility
changes sign. These qualitative differences are due to {\it exchange
interactions} between electrons near the Fermi surface and electrons
in the negative energy sea and to interband contributions to
Dirac electrons from charge and spin fluctuations.

Our aim in this work is to study the spin polarization dependence of quasiparticle properties in graphene particulary the renormalized velocity and inelastic scattering lifetime of quasiparticles within the leading-order single-loop self-energy expansion. Our theory for spin polarization dependence of quasiparticle velocity renormalization in interacting Dirac electron systems is motivated not only by fundamental many-body considerations, but also by application to improve high-speed operation in the spintronic devices~\cite{cho} and potential future experiments. By chemical doping in graphene, spin polarization effects are predicted for some adsorption configurations~\cite{zanella}. Remarkably, the studies of spin polarization dependence of quasiparticle properties should help to understand spin valve physics and recent measurements of electronic spin transport in garphene~\cite{tombros}, and the possibility of magnetism in graphene induced by single carbon atom defects \cite{magnetism}.

The paper is organized as the following. In Sec. II we introduce the formalism that will be used in calculating spin polarization quasiparticle properties which includes the many-body effects by suing RPA. In Sec. III we present our analytical and numerical results for the self-energy and renormalized Fermi velocity in doped graphene sheets. Sec. IV contains discussions and conclusions.

\section{METHOD AND THEORY }

We consider the long-range Coulomb electron-electron interaction.
We left out the intervalley scattering and use the two
component Dirac Fermion model. Accordingly, the total interacting Hamiltonian in
a continuum model at $K^+$ point is expressed as~\cite{slonczewski}

\begin{equation}\label{ham}
\hat{H}= -i\hbar
v\sum_i {\vec{\sigma}}\cdot{\bf \nabla}_i+
\frac{1}{2}\sum_{i\neq j}V ({\bf r}_i-{\bf r}_j),
\end{equation}

where $\vec{\sigma}$ are Pauli matrices and $v =3ta/2\hbar\simeq10^6$ m/s is the Fermi
velocity with $a\simeq1.42$ {\AA} is the carbon-carbon
distance in honeycomb lattice. Here ${\bf p}_i=-i\hbar{\bf \nabla}_i$ is the canonical momentum of the $i-$th electron and $v_q=2\pi e^2/\epsilon q$ is the
Fourier transform of the bare Coulomb interaction where $\epsilon$ is an average dielectric constant of the
surrounding medium. The coupling constant in graphene or
graphene's fine-structure constant
is $\alpha_{ee}=e^2/\epsilon\hbar v$. The coupling
constant in graphene depends only on the substrate dielectric
constant while in the conventional 2D electron systems is density
dependent. The typical value of dimensionless coupling constant is $0.25$ or $0.5$
for graphene supported on a substrate such a SiC or SiO$_2$.

As it is clearly seen from the first term of Eq. (\ref{ham}), the spectrum is
unbounded from below and it implies that the Hamiltonian has to be
accompanied by an ultraviolet cut-off which is defined $k_{\rm
c}$ and it should be assigned a value corresponding
to the wavevector range over which the continuum model
Eq.~(\ref{ham}) describes graphene. For definiteness we take
$k_{\rm c}$ to be such that $\pi k^2_{\rm c}=2(2\pi)^2/{\cal
A}_0$, where ${\cal A}_0=3\sqrt{3} a^2_0/2$ is the area of the
unit cell in the honeycomb lattice. With this choice, the energy $\hbar v k_{c}=7$ eV and
\begin{equation}
\Lambda =\frac{k_{\rm
c}}{k_{\rm F}}=\sqrt{\frac{2 g_{\rm v}}{n{\cal A}_0}}~.
\end{equation}
The continuum model is useful when $k_{\rm c} \gg k_{\rm F}$, {\it
i.e.} when $\Lambda \gg 1$. Note that, for instance, electron densities $n=0.36 \times 10^{12}$ and $0.36 \times 10^{14}$ cm$^{-2}$ correspond to $\Lambda=100$ and $10$, respectively.

The spin-polarization dependence of dynamical polarizability tensor in terms of one-body noninteracting Green's function is written as~\cite{Giuliani_and_Vignale}
\begin{eqnarray}\label{chi0}
&&\chi^{(0)}_{\sigma}({\bf q},\Omega,\mu)=-i\int \frac{d^2{\bf k}}{(2\pi)^2}
\int\frac{d\omega}{2\pi} \nonumber\\ && {\rm Tr}[i\gamma_0 G^{(0)}_{\sigma}({\bf k}+{\bf q},
\omega+\Omega,\mu) i\gamma_0 G^{(0)}_{\sigma}({\bf k},\omega,\mu)]
\end{eqnarray}

where $\sigma$ refers to the spin-direction, $\uparrow$ or $\downarrow$. After implementing $G^{(0)}_{\sigma}({\bf k},\omega,\mu)$ in Eq.(\ref{chi0})
and calculating the integral, the results end up to the
follow expression~\cite{yafis}
\begin{eqnarray}\label{eq:final_result}
&&\chi^{(0)}_{\sigma}({\bf q},i\Omega,\mu)=-g_v\frac{\mu^{\sigma}}{2\pi v^2}-
g_v \pi B/2\nonumber\\
&+&g_v B\Re e
\left[\arcsin(C)+C\sqrt{1-
C^2}\right]\,.
\end{eqnarray}
where $g_v=2$ is valley degeneracy. $\mu^{\sigma}$ is the spin dependence
chemical potential, $B=q^2/(8~\pi \sqrt{\Omega^2+v^2 q^2})$ and $C=(2\mu^{\sigma}+i\Omega)/(vq)$.

The technical calculation~\cite{Giuliani_and_Vignale} on which our conclusions are based is an evaluation of the spin-polarization dependence
electron self-energy $\Sigma^{\sigma}_s({\bf k},\omega)$ of the Dirac fermion near the quasiparticle-pole.
$\Sigma^{\sigma}_s({\bf k},\omega)$ describes the interaction of a single Dirac electron with spin $\sigma$ near the 2D Fermi surface with all states inside the Fermi sea, and with virtual particle-hole and collective excitations of the entire Fermi sea. A direct expansion of electron self-energy in powers of the Coulomb interaction is never possible in a 2D electron liquid because of the long-range of the Coulomb interaction. Our results for the Dirac electron gas are based on the
random phase approximation (RPA) in which the self-energy is expanded to the first order in the dynamically screened Coulomb interaction (setting $\hbar=1$):
\begin{eqnarray}\label{eq:sigma_rpa}
\label{eq:RPAse}
&\Sigma^{\sigma}_s&({\bf k},i\omega_n)=-\frac{1}{\beta}\sum_{s'}
\int \frac{d^2{\bf q}}{(2\pi)^2}\sum_{m=-\infty}^{+\infty}
\frac{v_q}{\varepsilon({\bf q},i\Omega_m,\zeta)}\nonumber\\
&\times&\left[\frac{1+s s'\cos{(\theta_{{\bf k},{\bf k}+{\bf q}})}}{2}\right]G^{0 \sigma}_{ s'}({\bf k}+{\bf q},i\omega_n+i\Omega_m)\,,
\end{eqnarray}
where $s=+$ for electron-doped systems and $s=-$ for
hole-doped systems, $\zeta$ is the spin polarization parameter, $\zeta=|n_{\uparrow}-n_{\downarrow}|/n$, $\beta=1/(k_{\rm B} T)$ and
$\varepsilon({\bf q},i\Omega_m,\zeta)$ is the RPA dielectric function. $n_{\sigma}$ is
the spin polarized density and $n$ is the total density of system. The RPA dielectric function is given by

\begin{equation}\label{eq:ex+corr}
\varepsilon({\bf q},i\Omega,\zeta)=1-v_q[\chi^{(0)}_{\uparrow}({\bf q},i\Omega,\zeta)+\chi^{(0)}_{\downarrow}({\bf q},i\Omega,\zeta)]~.
\end{equation}

In the Dirac 2D electron system, dielectric-function
contributions from intraband and interband excitations are subtly interrelated.  The two contributions must
be included on an equal footing in order to describe the Dirac fermion physics correctly. For example one key property, that the static dielectric function is independent of $q$ at small $q$, requires intraband and interband contributions to be summed.

For definiteness, we limit our discussion to
an electron-doped system with spin polarization dependence of chemical potential $\mu^{\sigma}$.

In Eq.~(\ref{eq:sigma_rpa}) $\omega_n=(2n+1)\pi/\beta$ is a fermionic Matsubara frequency, the sum runs over all the bosonic Matsubara frequencies $\Omega_m=2m\pi/\beta$.
The factor
in square brackets in Eq.~(\ref{eq:RPAse}), which depends on the angle $\theta_{{\bf k},{\bf k}+{\bf q}}$ between ${\bf k}$ and ${\bf k}+{\bf q}$, captures
the dependence of Coulomb scattering on the relative chirality $s s'$ of the interacting electrons.
The Green's function $ G^{0 \sigma}_{s}({\bf k},i\omega) = 1/[i\omega - \xi^{\sigma}_s({\bf k})]$ describes the free propagation of
states with wavevector ${\bf k}$, Dirac energy $ \xi^{\sigma}_s({\bf k})=sv k-\mu^{\sigma}$ (relative to the chemical potential) and chirality $s=\pm$.
The quasiparticle exciatation energy measured from the chemical potential can be given by Dyson equation $E^{\sigma}_{s} ({\bf k})=\xi^{\sigma}_{s} ({\bf k})+ \Re e \Sigma^{(\rm ret, \sigma)}_s({\bf k},\omega) $ evaluated at $\omega=E^{\sigma}_{s} ({\bf k})$. After continuation from imaginary to real frequencies, $i \omega \to
\omega + i \eta$ and using the Dyson equation, the spin dependence renormalized Fermi velocity can be expressed~\cite{Giuliani_and_Vignale} in terms of the wavevector and
frequency derivatives of the retarded self-energy $\Sigma^{(\rm ret, \sigma)}_+({\bf k},\omega)$ evaluated at the spin dependence Fermi surface which is $k_F^{\sigma}=(1+\sigma\zeta)^{1/2}k_F$, where $k_F$ is Fermi momentum:

\begin{equation}\label{eq:v_star_dyson}
\frac{v^{\star(Dyson)}_\sigma}{v}
=\frac{d E^{\sigma}_+({\bf k})}{d {\bf k}}=\frac{\displaystyle 1+(v)^{-1}\left.\partial_k \Re e \Sigma^{(\rm ret, \sigma)}_+({\bf k},\omega)\right|_{k=k^{\sigma}_{\rm F},\omega=0}}{1-\left.
\partial_{\omega} \Re e \Sigma^{(\rm ret, \sigma)}_+({\bf k},\omega)\right|_{k=k^{\sigma}_{\rm F},\omega=0}}\,.
\end{equation}

\begin{figure}[ht]
\begin{center}
\includegraphics[width=1.1\linewidth]{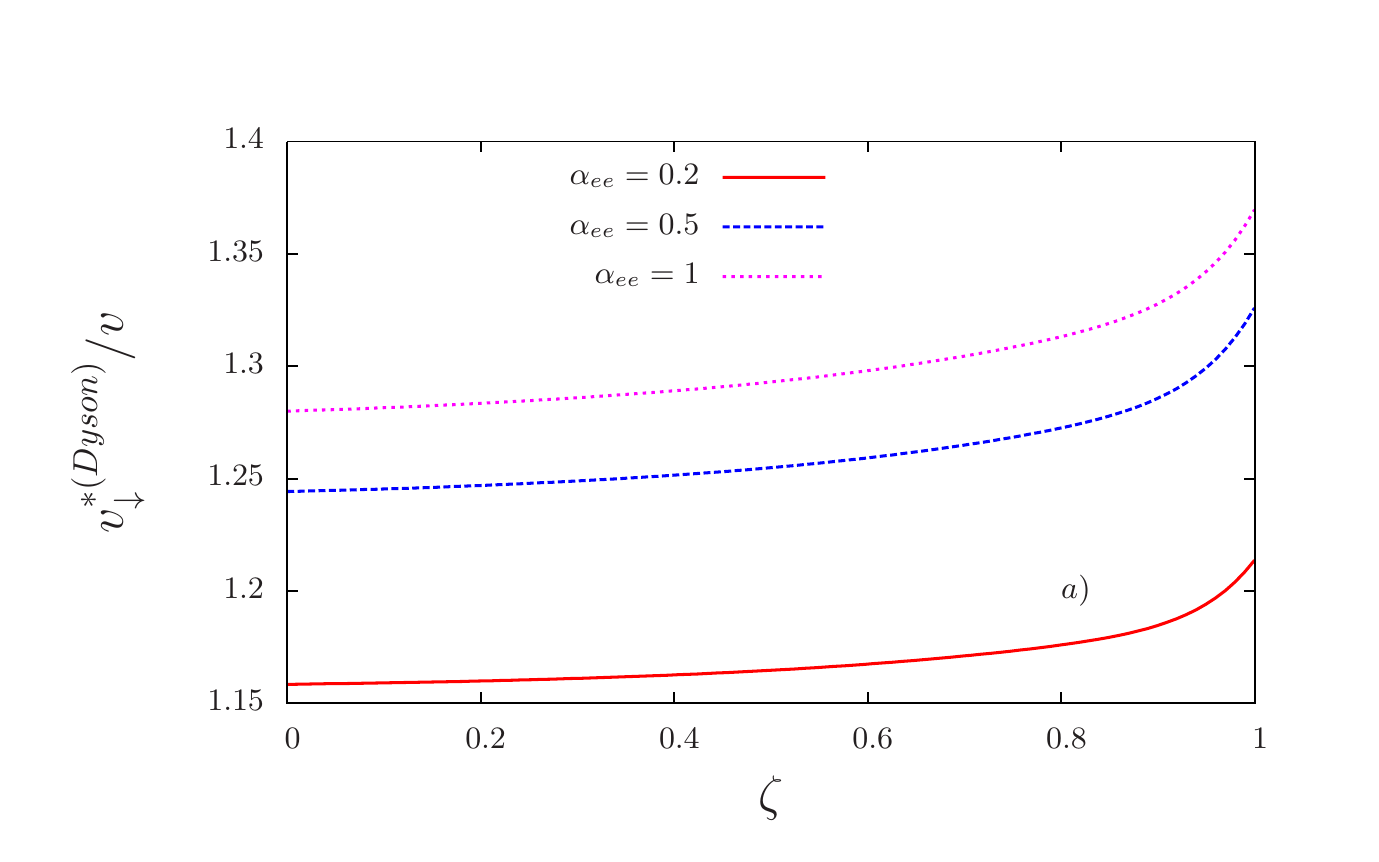}
\includegraphics[width=1.1\linewidth]{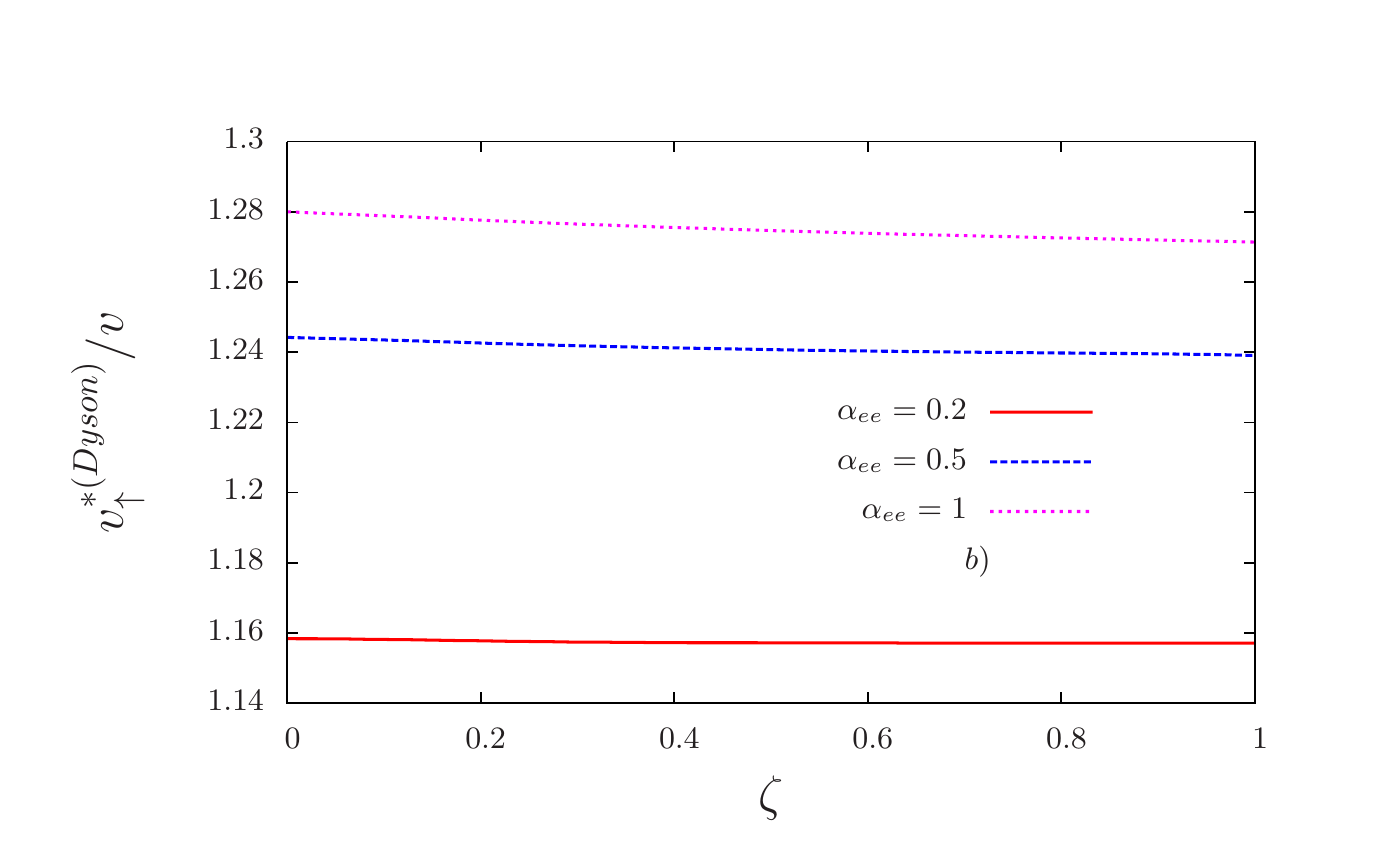}
\caption{(Color online) Spin-polarization dependence renormalized velocity, in the Dyson scheme, scaled by that of a noninteracting velocity as a function of the degree of spin polarization $\zeta$ for cut-off value $\Lambda=100$ ($n=0.36 \times 10^{12}$ cm$^{-2}$) in (a) the down-spin where $0 < n_{\downarrow}<0.18\times 10^{12} $ cm$^{-2}$  and (b) the up-spin where $ 0.18\times 10^{12} < n_{\uparrow}<0.36\times 10^{12} $ cm$^{-2}$  for different coupling constant values. The spin polarization dependence of the up- and down-spin velocity behaves differently as $\zeta$ increases.}
\end{center}
\end{figure}

In the on-shell approximation ( OSA), on the other hand, the renormalized
velocity is given by
\begin{eqnarray}\label{eq:v_star_osa}
\frac{v^{\star(OSA)}_\sigma}{v} &&=1+(
v)^{-1}\partial_k\Re e[ \Sigma_+^{(ret,\sigma)}({\bf
k},\omega)]|_{\omega=0,k=k^{\sigma}_{\rm F}}\nonumber\\
&&+
\partial_{\omega}\Re e[ \Sigma_+^{(ret,\sigma)}({\bf
k},\omega)]|_{\omega=0,k=k^{\sigma}_{\rm F}}
\end{eqnarray}

This expression can also be obtained from the formal definition of $v^*$ given in the first equality in Eq.~(\ref{eq:v_star_dyson}) when the second term in the Dyson equation, $\Re e \Sigma^{(\rm ret, \sigma)}_+({\bf k},\omega) $, is evaluated at the bare pole, $\omega=\xi^{\sigma}_+ ({\bf k})$. The OSA thus gives the quasiparticle velocity to the first order in the retarded self-energy. The renormalized velocity in
this approximation demonstrates qualitatively the same behavior
obtained by the Dyson equation, Eq.~(\ref{eq:v_star_dyson}) but its
magnitude is larger than the one calculated within the Dyson
scheme.

The quasiparticle weight factor $Z_{\sigma}$ evaluated at the spin dependence Fermi surface and given
by $Z_{\sigma}^{-1}=1-\left.
\partial_{\omega} \Re e \Sigma^{(\rm ret, \sigma)}_+({\bf k},\omega)\right|_{k=k^{\sigma}_{\rm F},\omega=0}$. In the up-spin case, the majority-spin, $Z_{\uparrow}$ value is a bit smaller and the down-spin case, the minority-spin, $Z_{\downarrow}$ value is bigger than the results of $Z(\zeta=0)$.

\section{NUMERICAL RESULTS}

Since the single-particle self-energy, the density of
states, the dynamical screening, the Fermi momentum,
and the Fermi energy in the chiral Dirac fermion are all affected by spin polarization, we expect all Fermi liquid parameters to be strongly dependent on the spin-polarization
parameter. An important thermodynamic quantity is the system
compressibility, which has been already studied by two of us~\cite{alireza}.

Our results for spin-polarization dependence of the Dirac electron velocity, ${v^\star}_\sigma/v$, at fixed electron density value in the up- and down-spin, the majority- and minority-spin as a function of the $\zeta$ are summarized in Fig.~1 for different values of the dimensionless coupling constant, $\alpha_{ee}$. For the up-spin Dirac electron, renormalized velocity decreases with increasing spin-polarization degree of freedom. However, the down-spin electron renormalized velocity increases by increasing spin-polarization. These behaviors are based on the effect of the exchange energy in the spin channels between electrons near the Fermi surface. Since the electron density in the down-spin channel, $n_{\downarrow}=n(1-\zeta)/2$ is less than the electron density in the up-spin channel $n_{\uparrow}=n(1+\zeta)/2$, and $n_{\downarrow}$ decreases however $n_{\uparrow}$ increases by increasing $\zeta$, therefore, the exchange contribution of the down-spin is dominated and results in increasing the renormalized velocity in the spin-down channel and decreasing the renormalized velocity in the spin up-channel. In contrast, the 2DEG where the down-spin mass increases with spin-polarization first and as $\zeta$ approaches near to one, it decreases sharply, the spin-polarized down-spin velocity tends to a constant when $\zeta$ reaches to unity~\cite{zhang}. It should be noticed that the up- and down-spin Fermi velocities are the same value at $\zeta=0$.

\begin{figure}[ht]
\begin{center}
\includegraphics[width=1.1\linewidth]{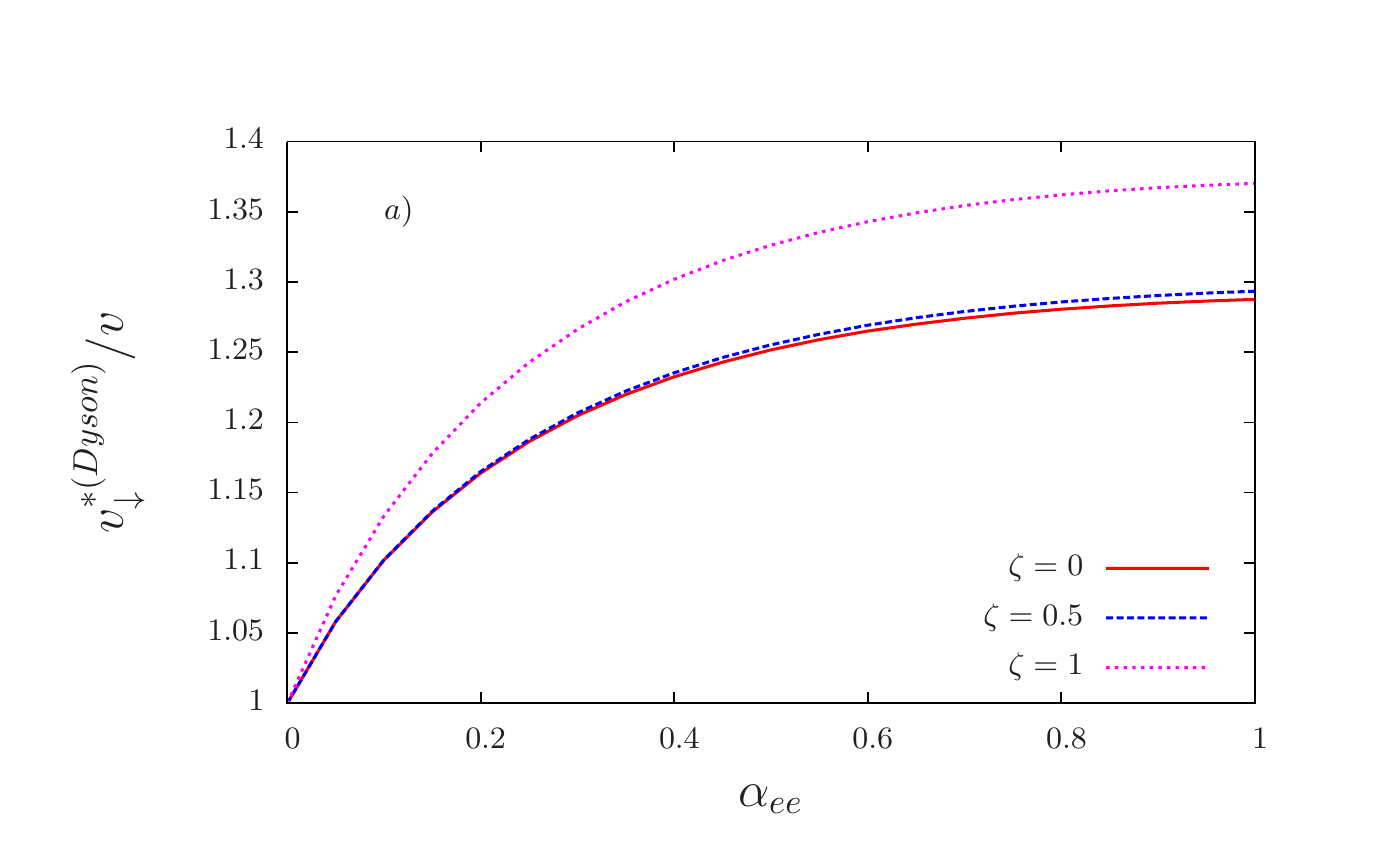}
\includegraphics[width=1.1\linewidth]{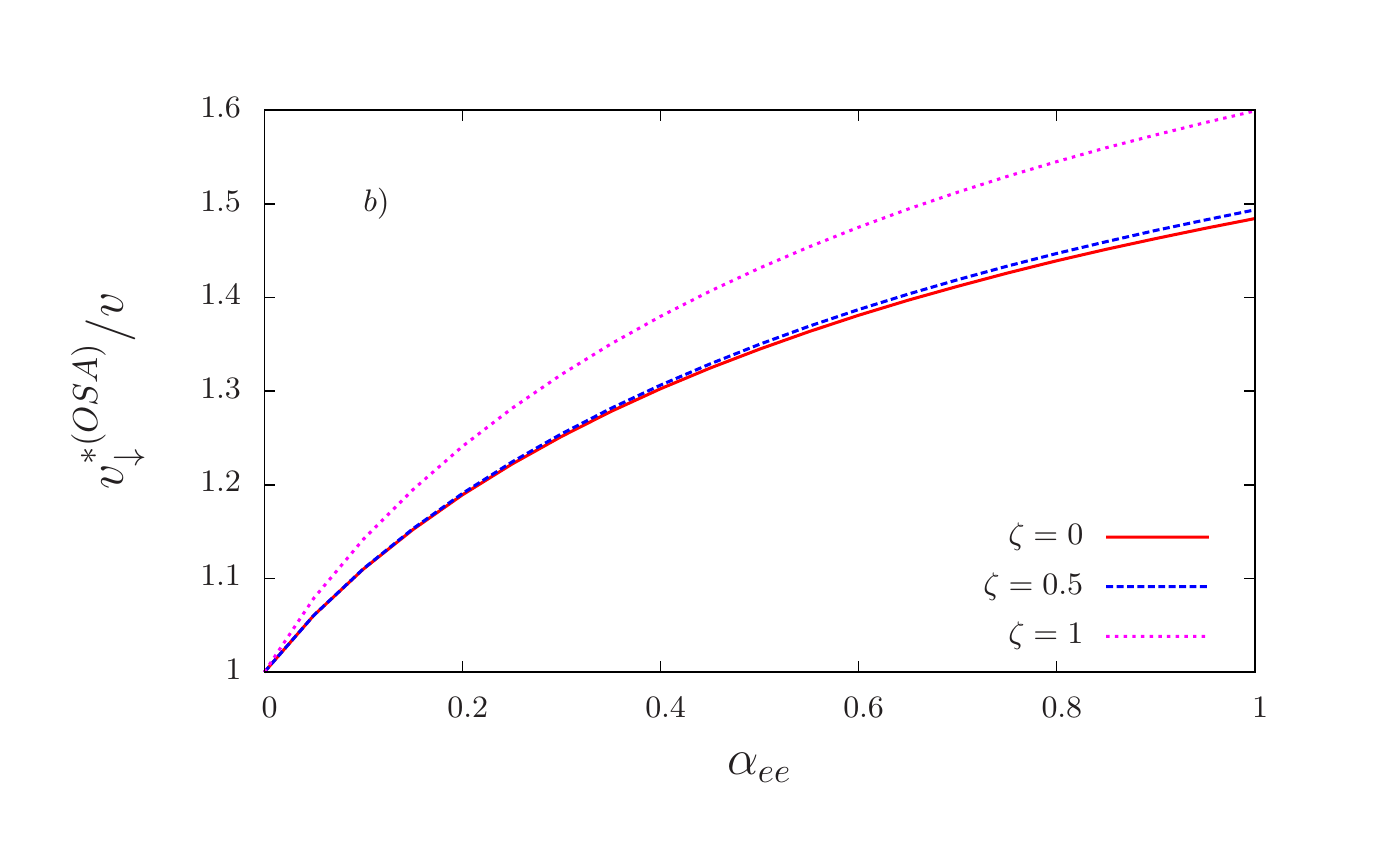}
\caption{(Color online) Renormalized velocity of the down-spin scaled by that of a noninteracting velocity, $v^*_{\downarrow}/v$, as a function of the coupling constant $\alpha_{ee}$ for cut-off value $\Lambda=100$ ($n=0.36 \times 10^{12}$ cm$^{-2}$) in (a) Dyson and (b) OSA approximations given by Eqs. (\ref{eq:v_star_dyson}) and (\ref{eq:v_star_osa}), respectively.}
\end{center}
\end{figure}

\begin{figure}[ht]
\begin{center}
\includegraphics[width=1.1\linewidth]{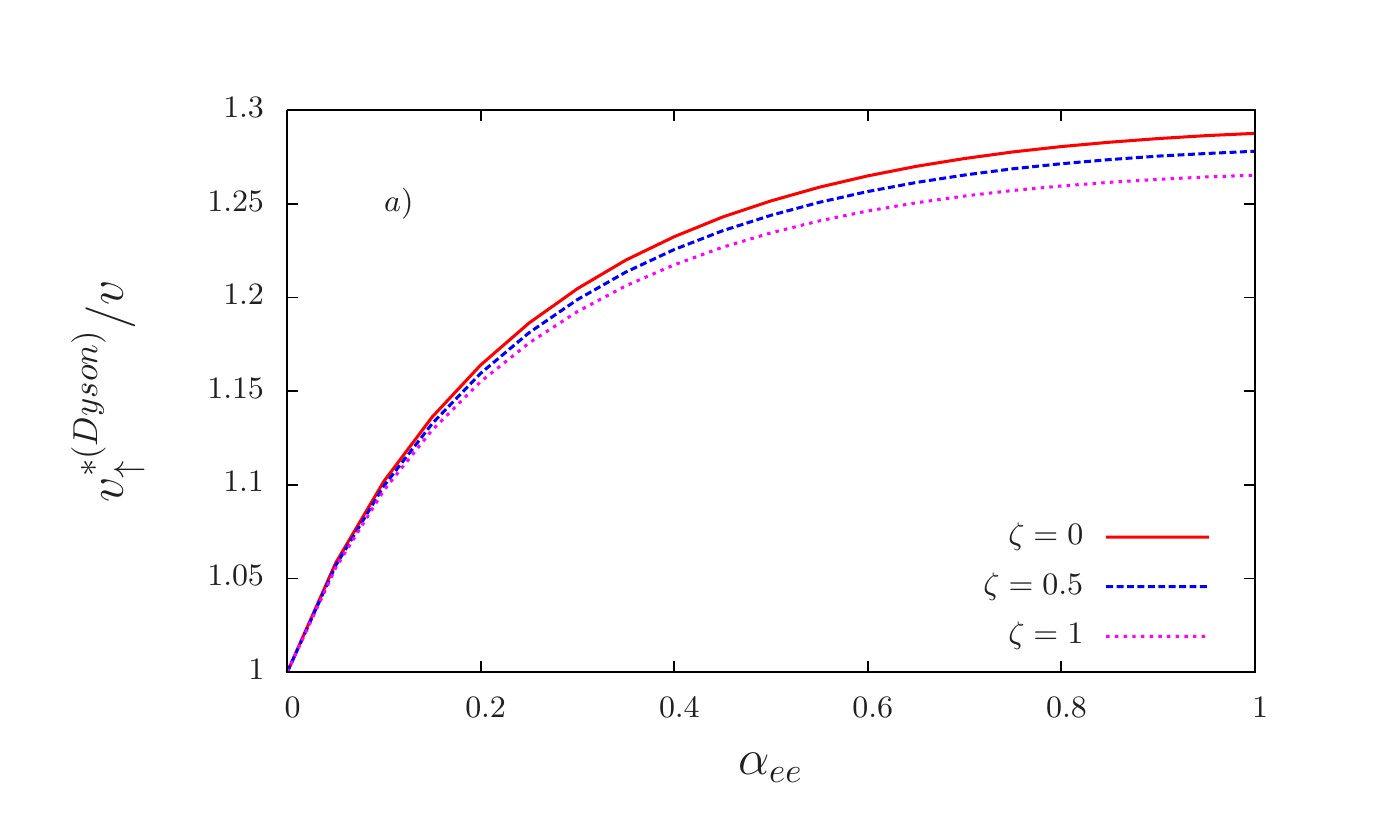}
\includegraphics[width=1.1\linewidth]{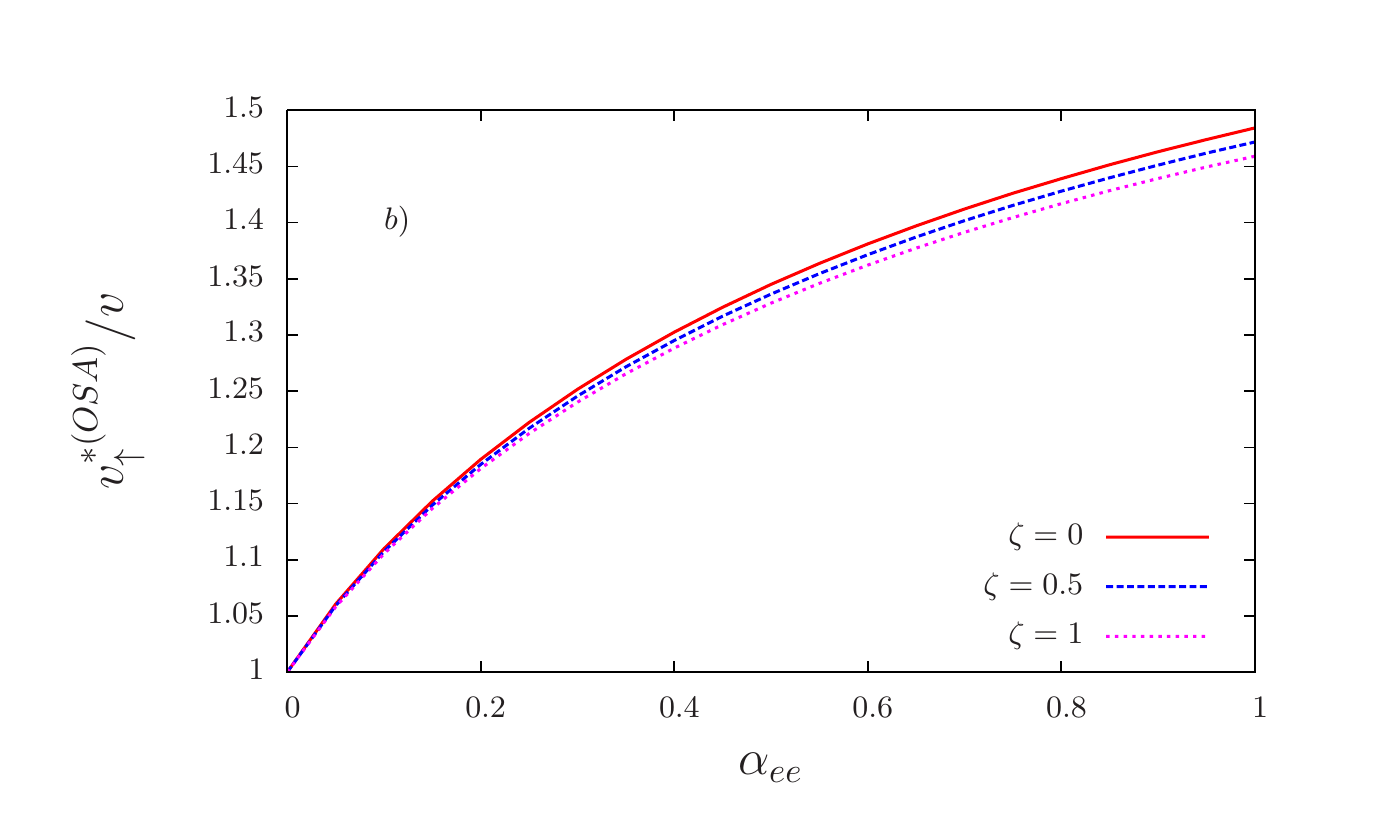}
\caption{(Color online) Renormalized velocity of the up-spin scaled by that of a noninteracting velocity, $v^*_{\uparrow}/v$, as a function of the coupling constant $\alpha_{ee}$ for cut-off value $\Lambda=100$ ($n=0.36 \times 10^{12}$ cm$^{-2}$) in (a) Dyson and (b) OSA approximations given by Eqs. (\ref{eq:v_star_dyson}) and (\ref{eq:v_star_osa}), respectively.}
\end{center}
\end{figure}

As it has been discussed previously~\cite{yafis, polini, polini2},
graphene's Fermi liquid properties depend only weakly on the carrier
density which is expressed in terms of the cut-off parameter
$\Lambda$. The trends exhibited in Fig.~1 can be understood by
considering the limits of small $\alpha_{ee}$ and the limit of large
$q$ at all values of $\alpha_{ee}$. In the former limit screening is
weak except at extremely small $q$. The self-energy can be decomposed as the sum of a contribution from the interaction of quasiparticles
at the Fermi energy, the {\em residue} contribution $\Sigma^{(\rm res, \sigma)}$, and a contribution from interactions with quasiparticles
far from the Fermi energy and via both exchange and virtual fluctuations, the {\em line} contribution $\Sigma^{(\rm line, \sigma)}$. In $\partial_{\omega}
\Sigma^{(\rm res, \sigma)}_+({\bf k},\omega)$, for example, the
integral over $q$ diverges logarithmically at small $q$ when
$\varepsilon({\bf q},\omega=0)$ is set equal to one, {\em i.e.} when
screening is neglected. Accordingly, screening cuts off this logarithmic
divergence at a wavevector. More precisely, we have  
\begin{eqnarray}
&&\frac{\partial}{\partial {\omega}} \Sigma^{(\rm res, \sigma)}_+({\bf k},\omega)|_{k=k_F^{\sigma}, \omega=0}=\frac{\alpha_{ee}}{2 \pi }\int_0^{2\sqrt{1+\sigma \zeta}}dx\nonumber\\
&&\frac{1}{x~\varepsilon(x,0)}\frac{4-x^2/(1+\sigma\zeta)}{\sqrt{4-x^2/(1+\sigma\zeta)}}
\end{eqnarray}

and notice that $\partial_\omega\Re e
\Sigma^{(\rm res, \sigma)}_+(k_{\rm F},0)=0$ for a case that $\sigma\zeta=-1$.
Because $\varepsilon({\bf q},\omega=0)$ happens to be
independent of $q$ for transitions between Fermi surface points, it
is possible to evaluate $\partial_{\omega} \Sigma^{(\rm res,
\sigma)}_+({\bf k},\omega)$ for the case that $\sigma\zeta\neq-1$ analytically. We find that
\begin{eqnarray}\label{eq:partial_omega_residue_final_4}
\frac{\partial}{\partial \omega}\Re e \Sigma^{(\rm res, \sigma)}_+(k_{\rm F},\omega=0)=\frac{\alpha_{ee}}{2 \pi}\times\nonumber\\
\left[\sqrt{4-{\eta}_{\sigma}^2}\ln{\left(\frac{2+\sqrt{4-{\eta}_{\sigma}^2}}{{\eta}_{\sigma}}\right)}-\frac{1}{2}(4-{\eta}_{\sigma}\pi)\right]\,,
\end{eqnarray}
where
${\eta}_{\sigma}=g_v\alpha_{ee}
(\sqrt{1+\zeta}+\sqrt{1-\zeta})/\sqrt{1+\sigma\zeta}$. It should be worthwhile mentioning that for $\eta_{\sigma}>2$ we use an equality in which $-i\ln(x+i\sqrt{1-x^2})=\arccos(x)$ when $x\leq 1$. Importantly, similar $\zeta$ dependence appears in the $v^*_{\sigma}$
at small $\alpha_{ee}$. More precisely, the $\zeta$ dependence of
$v^*_{\sigma}(\zeta)$ , both in the Dyson and OSA schemes, in the
$\alpha_{ee}\rightarrow 0$ limit, for $\zeta <1$, is given
by

\begin{equation}\label{eq:velocity_limit}
\frac{v^*_{\sigma}(\zeta)}{v}-1=\frac{\alpha_{ee}}{\pi}\left[ \ln(g_v\alpha_{ee})+ \ln(\frac{\eta_{\sigma}}{2\alpha_{ee}})\right]~.
\end{equation}

This analytical expression shows that the renormalized velocity in the down-spin enhances while it decreases in the up-spin channel. In the limit of small $\zeta$, Eq.~(\ref{eq:velocity_limit}) is simplified and $v^*_{\uparrow \downarrow}(\zeta)/v-1=\alpha_{ee}\left[ \ln(g_v\alpha_{ee})\mp \zeta/2\right]/\pi$. All these behaviors are very familiar from the case of the effective mass or the effective Fermi velocity in a normal 2DES but more significantly, the spin-polarization term is different than that the 2DEG~\cite{zhang, note}. The discrepancy is due to the nature of the chiral Dirac electron behavior in graphene flake with having the linear dispersion relation.

\begin{figure}[ht]
\begin{center}
\includegraphics[width=1.1\linewidth]{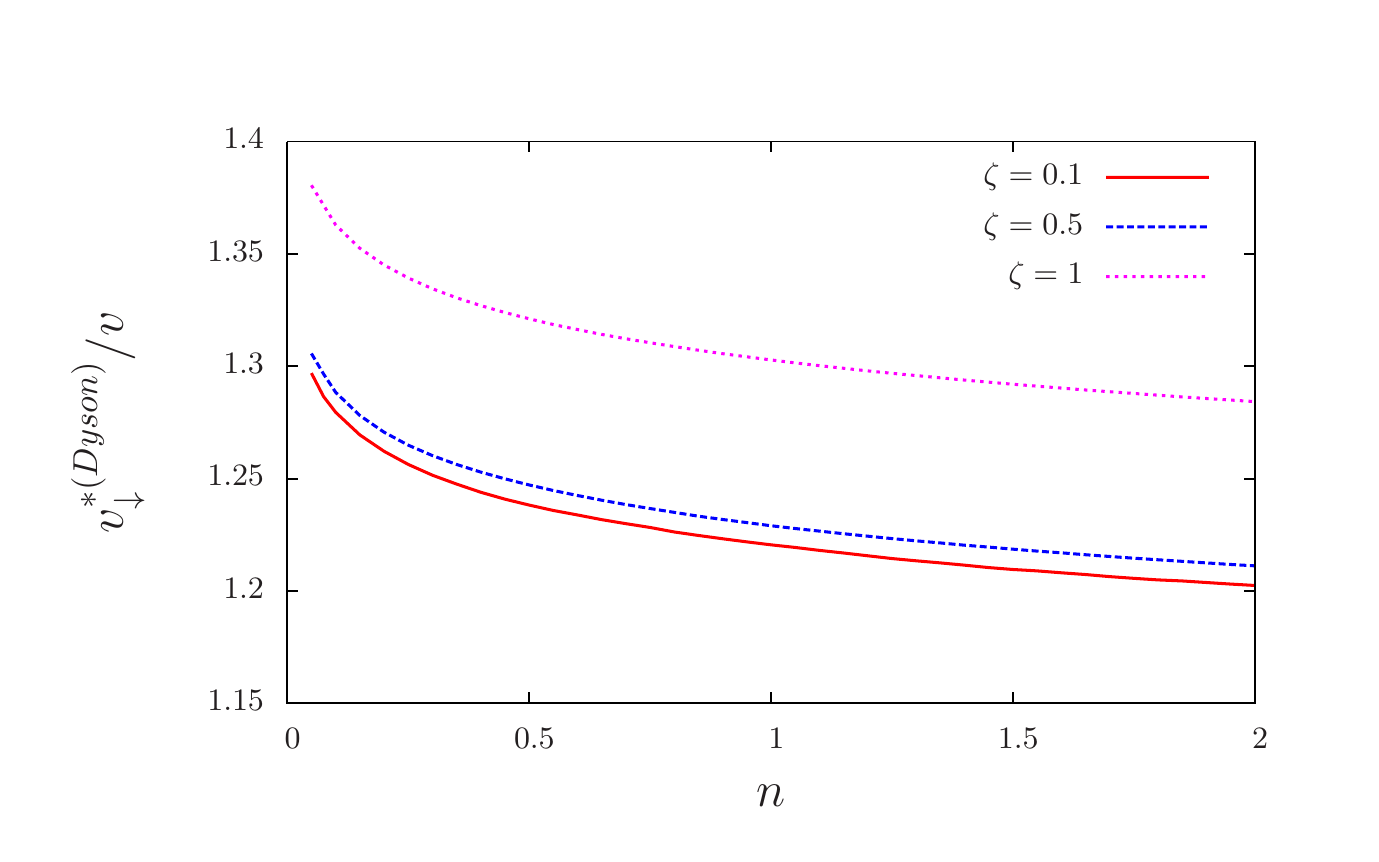}
\caption{(Color online) Renormalized velocity of the down-spin, in the Dyson scheme,  scaled by that of a noninteracting velocity as a function of the electron density
(in units of $10^{12}~{\rm cm}^{-2}$) for different $\zeta$ values at $\alpha_{ee}=0.5$.}
\end{center}
\end{figure}

In Fig.~2, we show the down-spin renormalized velocity scaled by that of a noninteracting velocity as a function of the coupling constant in both the dyson and OSA approximations which are defined by Eqs.~(\ref{eq:v_star_dyson}) and (\ref{eq:v_star_osa}), respectively. Clearly, the velocity values increase significantly when $\zeta$ approaches to unity. Despite the strong down-spin velocity dependence of the spin degree of freedom, the up-spin velocity becomes smoothly smaller with $\zeta$ as it shown in Fig.~3. Notice that $v_{\sigma}^{* (OSA)}$ is always larger than $v_{\sigma}^{* (Dyson)}$. Moreover, the $\zeta$ dependence of the renormalized velocity is opposite with respect to the spin direction. It would be worthwhile finding the asymptotic behavior of $v^*_{\sigma}$ at some conditions. At large $q$, interband charge fluctuations
dominate $\varepsilon({\bf q},\omega)-1$, which approaches
its simple undoped system form. It becomes especially clear when $\omega$ is expressed in units of $vq$ that the
typical value of $\varepsilon({\bf q},\omega)$ at large $q$ is $\sim 1$ with a non-trivial dependence on $\alpha_{ee}$.
The $q$ integrals all vary as $q^{-1}$, requiring that the Dirac electrons model be accompanied by an ultraviolet
cut-off. Since the crossover between intraband and interband screening occurs for
$q \sim k_{\rm F}$, it follows that $\partial_k \Sigma^{\rm res}$ and
$\partial_\omega \Sigma^{\rm res}$ have contributions that are analytic in $\alpha_{ee}$ and vary as $\ln(\Lambda)$ when $\Lambda$ is large. To leading order in $\ln(\Lambda)$ we find that
$v^\star/v- 1 = \alpha_{ee}[1-2g_v\alpha_{ee}g(2g_v\alpha_{ee})]\ln{(\Lambda)}/4$ which is $\sigma$ independent and $g(x)$ is defined in Ref.[\onlinecite{polini}].

\begin{figure}[ht]
\begin{center}
\includegraphics[width=1.1\linewidth]{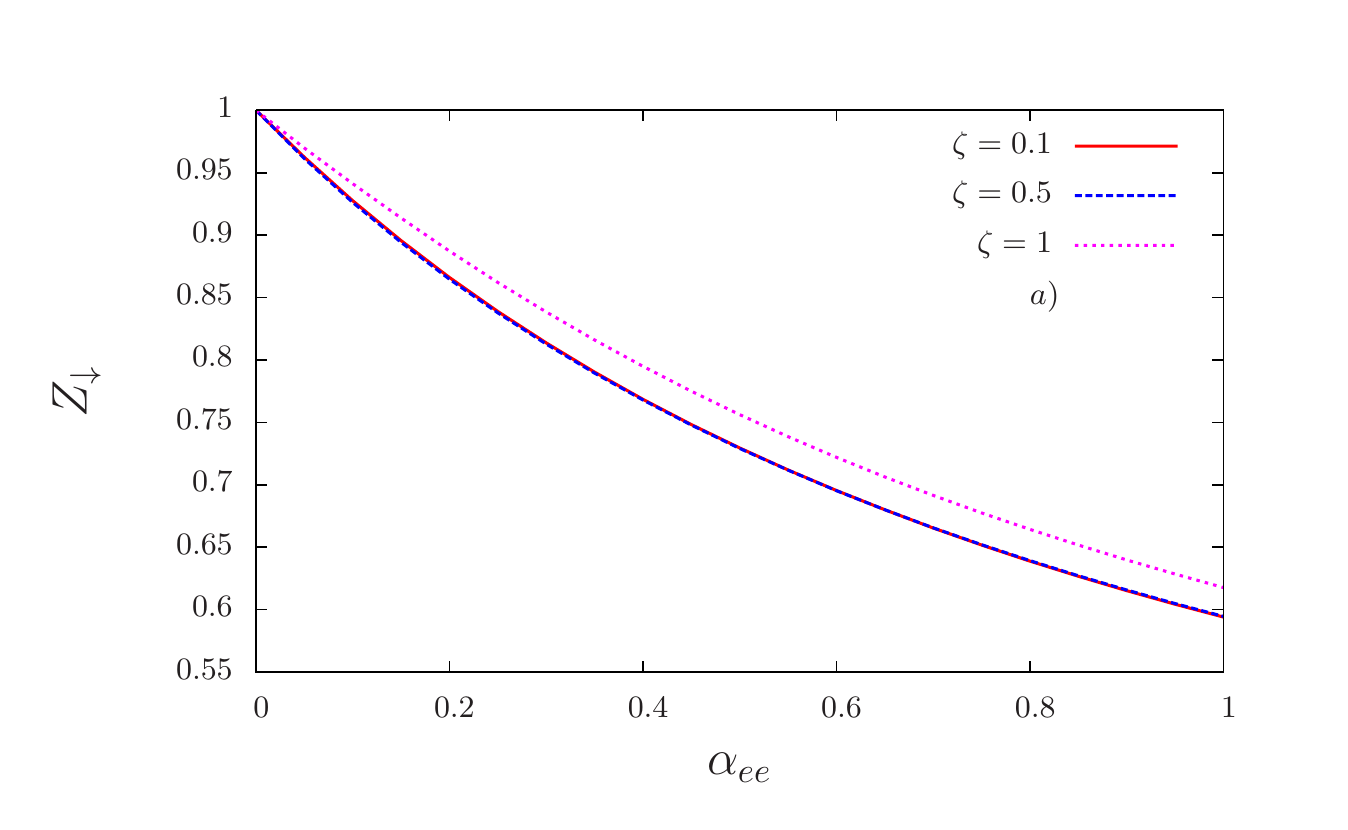}
\includegraphics[width=1.1\linewidth]{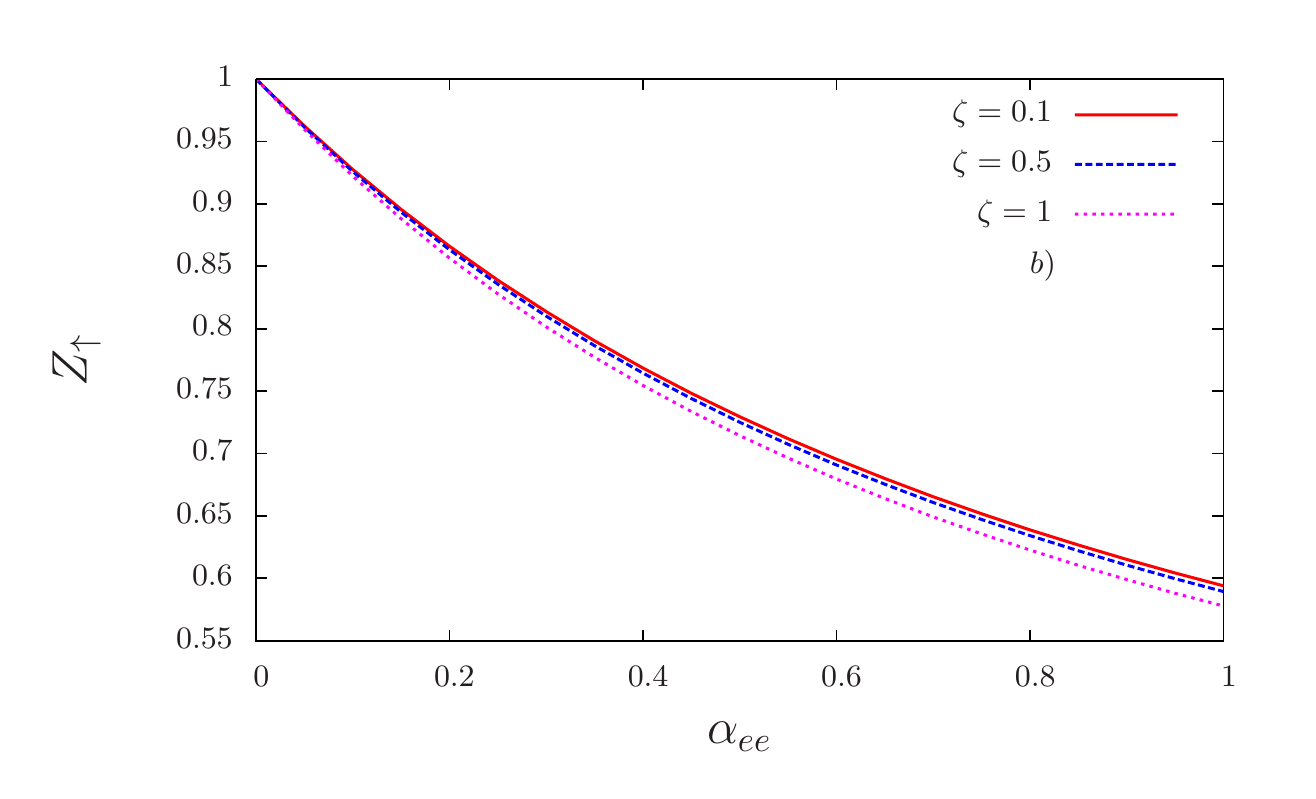}
\caption{(Color online) Renormalized constant $Z_{\sigma}$ for (a) the down-spin and (b) the up-spin as a function of the coupling constant $\alpha_{ee}$ for cut-off value $\Lambda=100$ ($n=0.36 \times 10^{12}$ cm$^{-2}$).}
\end{center}
\end{figure}

In Fig.~4 we show the down-spin renormalized velocity as a function of the electron density (in units of $10^{12}$ cm$^2$). In contrast to the 2DEG, the renormalized velocity increases by decreasing the electron density and indicates no Wigner crystallization~\cite{Giuliani_and_Vignale} occurs in pristine Dirac fermion systems~\cite{dahal}. Note that at very small $n$, the system is highly correlated and a model going beyond the RPA is necessary to account for increasing correlation effects
at low density~\cite{sarma}. Our theoretical calculations show that, even at moderately low densities, the velocity enhancement in a supported graphene sheet can vary a lot in qualitatively good agreement with measured data in a suspended graphene sheet~\cite{elias}.

\begin{figure}[ht]
\begin{center}
\includegraphics[width=1.1\linewidth]{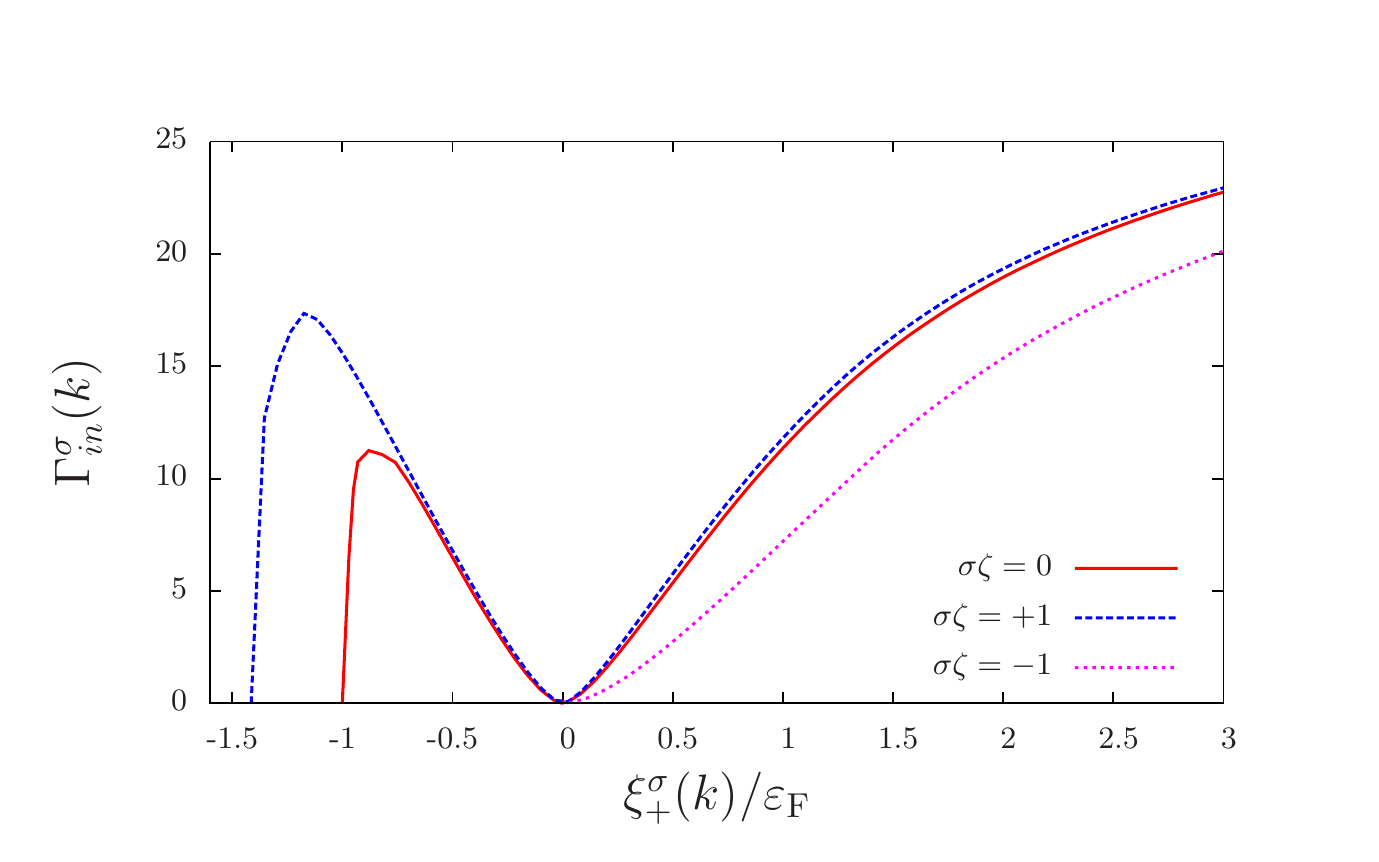}
\caption{(Color online) Inverse inelastic
scattering lifetime of quasiparticle, $\Gamma_{in}^{\sigma}(k)$ (in units of psec$^{-1}$) in graphene as a function of the
on-shell energy, $\xi_+^{\sigma}(k)$ for different spin polarization values. The data in this figure refer to $n=5\times10^{12}$cm$^{-2}$ and $\alpha_{ee}=0.25$.}
\end{center}
\end{figure}

We have also calculated the
renormalization factor $Z_{\sigma}(\alpha_{ee}, \zeta)$ which is equal to the
discontinuity in the momentum distribution at $k^{\sigma}_F$.
The effect of $\zeta$ is to make the $Z_{\downarrow}$ values
larger at large $\alpha_{ee}$ compared to the case when $\zeta$ is not included as shown in Fig.~5. The non-zero values of $Z_{\sigma}$, shows the Fermi liquid picture
in the whole range of $\alpha_{ee}$ and $\zeta$. When $n\rightarrow 0$ the $Z_{\sigma}$ factor drops to zero logarithmically. Notice that in leading order of $\ln{\Lambda}$, the renormalization factor is independent of $\sigma$ and behaves like $Z^{-1}- 1 = \alpha_{ee} \lambda(2 g_v\alpha_{ee})\ln{(\Lambda)}/6$ where $\lambda(x)$ is defined in Ref.[\onlinecite{polini}].

Finally, we compute the inelastic scattering lifetime of quasiparticles due to
carriers-carriers interactions at zero temperature for different $\zeta$ values. This is obtained through the imaginary part of the self-energy when
the frequency evaluated at the on-shell energy
\begin{eqnarray}
{\tau_{in}^{\sigma}}^{-1}({\bf k})=\Gamma_{in}^{\sigma}({\bf k})=-\frac{2}{\hbar}\Im m \Sigma_+^{(\rm ret,\sigma)}({\bf k},\xi_+^{\sigma}({\bf
k}/\hbar)),
\end{eqnarray}
where $\Gamma_{in}^{\sigma}({\bf k})$ is the quantum
level broadening of the momentum with eigenstate $|{\bf k}>$. It is worthwhile to note that
the expression of ${\tau_{in}^{\sigma}}^{-1}({\bf k})$ is identical with a result obtained by the Fermi's golden rule summing the scattering rate of electron and hole contributions at wave vector ${\bf k}$~\cite{Giuliani_and_Vignale}. Fig.~6 shows the behavior of the spin polarization dependence of the inverse inelastic scattering lifetime for
$n=5\times10^{12}$cm$^{-2}$ and $\alpha_{ee}=0.25$. Imaginary
part of the self-energy evaluated at the on-shell energy starts
from $\xi_+^{\sigma}(k)=-\varepsilon_{\rm F}^{\sigma}$, exhibits a minimum at zero energy and then grows up. Scattering rate in graphene is a
smooth function which is in contrast with the conventional 2D
semiconductors and 2D electron liquids because of the absence of
both plasmon emission and interband processes~\cite{inelastic}. We also see in Fig.~6 that the scattering rate is quite sensitive to the spin polarization and the inelastic lifetime for minority spins is larger evidently than the
majority spin lifetime. The ratio of
the majority- to minority-spin lifetime is smaller than unity and related directly
to the polarization and electron energy.

\section{Conclusions}

In summary we have calculated the spin-polarized dependence of the quasiparticle in gaphene sheets and they could be strongly spin-polarization dependent and
substantially different than the usual unpolarized paramagnetic
values. Similar to a two-dimensional paramagnetic diluted magnetic semiconductor
electron gas~\cite{chang}, the Dirac electron Fermi velocity is highly spin dependent even if the spin polarization of the carrier population is negligibly small. Therefore, the spin-polarization dependence of chiral carrier transports can be observed experimentally nearly full spin polarization regimes. The majority-spin electron renormalized velocity decreases with increasing spin-polarization degree of freedom. However, the minority-spin electron renormalized velocity increases by increasing spin-polarization due to reduction in the electron density and consequently increase in the interaction between electrons near the Fermi surface. We show that the ratio of the lifetimes of majority- to minority-spin electrons is smaller than unity and related to the polarization and electron energy. It has important implications for the interpretation of many
types of spin polarized experiments. Our results might be used in calculating the effective density of states in graphene spintronic systems. The spin-polarized features that are the subject of this work may, in the future, lead to the development of graphene devices incorporating interference-based spin filters.

\section{Acknowledgement}

We would like to thank M. Polini for useful discussions. A. Q. was supported by EU-ICT-7 contract No. 257159 "MACALO".

\vspace{0.6cm}
\noindent




\begin{thebibliography}{99}
\bibitem{note}
Note that the effective mass in 2DEG at small $r_s$ and $\zeta$ values behaves like $m^*_{\uparrow \downarrow}/m=1+(r_s/\sqrt{2}\pi)\ln{r_s}\mp(r_s \zeta/2\sqrt{2} \pi)\ln{r_s}$~.
\end{thebibliography}
\end{document}